\documentclass[useAMS]{mn2e}

\usepackage{graphicx,amssymb,amstext} 

\begin{document}

\title[A halo model of Local IRAS galaxies using CLFs]{A Halo Model of Local {\it IRAS} Galaxies Selected at 60 $\mu$m Using Conditional Luminosity Functions} 
\author[L. Wang et al.]{
\parbox[t]{\textwidth}{
Lingyu Wang$^{1}$\thanks{E-mail: lingyu.wang@sussex.ac.uk}, Asantha Cooray$^{2}$, Seb Oliver$^{1}$}
\\
$^{1}$Astronomy Centre, Department of Physics and Astronomy, University of Sussex, Falmer, Brighton, BN1 9QH\\
$^{2}$Department of Physics \& Astronomy, University of California, Irvine, CA 92697, USA\\
}

\date{Accepted . Received ; in original form }

\maketitle

\begin{abstract}
Using conditional luminosity functions (CLFs) which encode the luminosity distribution of galaxies as a function of halo mass, we construct a halo model of IRAS galaxies selected at 60 $\mu$m. An abundance matching technique is used to link galaxy luminosity to the host halo mass. The shape of the mass - light relation at 60 $\mu$m is different from those derived at r-, K- and B-band. This is because the 60 $\mu$m LF can not be fitted by a Schechter function with a sharp exponential cutoff above $L_*$. We then seek the parameters in the CLFs that best fit the 60 $\mu$m LF and power spectrum. We find that the predicted galaxy bias as a function of $L_{60}$ from the best-fit model agrees well with the clustering measurement in different luminosity bins. At the faint end of the LF where quiescent star-forming galaxies dominate, most IRAS galaxies are central galaxies in halos of $M\gtrsim 10^{10}~h^{-1}~M_{\odot}$ but a non-negligible fraction are satellites typically hosted in more massive halos. The majority of IRAS galaxies with $L_{60} \gtrsim 10^{10}~h^{-2}~L_\odot$ are M82 type starbursts which are central galaxies hosted in halos of $M\gtrsim 10^{12.5}~h^{-1}~M_{\odot}$. In comparison, optical galaxies generally reside in much more massive halos. The rate of change in $L_{60}$ (an indicator of recent star formation) as a function of halo mass at $M\gtrsim 10^{12.5}~h^{-1}~M_{\odot}$ is much larger that $d L_{\rm optical} / dM$ or $d L_{\rm NIR}/dM$ indicating the existence of physical mechanisms which are very efficient in converting cold gas into stars, possibly dynamical effects arising from interactions or mergers. We further calculate the space density of major mergers for halos massive enough to host ultraluminous infrared galaxies (ULIRGs) using the mean merger rate derived from the Millennium simulations. Compared to the space density of local ULIRGs, it implies that either the majority of major mergers at $z\sim0$ do not lead to ULIRGs or the ULIRG phase is relatively short.
\end{abstract}


\section{Introduction}

Understanding the relative distribution between different populations of galaxies and the underlying dark matter is one of the most important goals in modern astrophysics and cosmology. In the local Universe, large galaxy spectroscopic surveys covering thousands of square degrees (e.g. SDSS and 2dFGRS) reveal how galaxies are distributed spatially and how that varies with intrinsic galaxy properties such as colour, luminosity and spectral type. A lot of progress have also been made in mapping galaxy distribution in the distant Universe albeit over much smaller areas (e.g. COSMOS and DEEP2). On the other hand, numerical simulations give us a fairly good understanding of the abundance and clustering of dark matter haloes. There are in general three different approaches to connect galaxies with dark matter halos. Hydro-dynamical simulations attempt to predict galaxy formation and evolution from first principles but it is computationally too expensive to carry out cosmological simulations with enough resolution and volume (e.g. Katz et al. 1992; Evrard et al. 1994; Frenk et al. 1996; Weinberg et al. 1997; Springel \& Hernquist 2003; Springel 2005; for a review see Springel 2010). In the foreseeable future, it seems infeasible to use hydro-dynamical simulations to explore a reasonable range of parameter space related to baryonic physics. Semi-analytic models (SAMs) also attempt to simulate the formation and evolution of galaxies in an {\it a priori} fashion but with highly simplified analytic prescriptions to approximate star formation and feedback processes (e.g. Kauffmann et al. 1999; Somerville \& Primack 1999; Benson et al. 2000, 2001; Springel et al. 2005; Somerville et al. 2008). It is computationally inexpensive which means that it is possible to use MCMC techniques to find the best-fitting parameters to reproduce observables such as galaxy luminosity function (LF), mass function and clustering. However, there are many degeneracies between various parameters in SAMs. More importantly, many physical processes (such as different feedback mechanisms) are still poorly understood which results in large uncertainties in various recipes used to follow galaxy formation and evolution. In recent years, an empirical approach based on virialized halos has been developed to circumvent these physical uncertainties. The basic halo model of the large-scale distribution of galaxies is a statistical approach that links galaxies with their host dark matter halos (e.g. Jing et al. 1998, 2002; Peacock \& Smith 2000; Seljak 2000; Scoccimarro et al. 2001; Berlind \& Weinberg 2002; Cooray \& Sheth 2002; Zheng 2004; Zehavi et al. 2004, 2005; Collister \& Lahav 2005; Tinker et al. 2005; Skibba et al. 2007; Brown et al. 2008). The halo occupation distribution (HOD) gives the the number of galaxies of certain intrinsic properties (such as luminosity, colour) in a halo of mass $M$. Combined with an assumption about the spatial distribution of these galaxies in the halo, the abundance and clustering of haloes can be used to derive those of galaxies. One of the disadvantages of the halo model is that it does not directly provide a physical understanding of the resulting halo occupation function. 

In this paper, we apply an important extension of the basic halo model, the so-called conditional luminosity function (CLF), to local IRAS galaxies selected at 60 $\mu$m. CLF $\Phi(L|M)$ describes the number of galaxies as a function of luminosity and halo mass (Yang et al. 2003, 2005; Cooray \& Milosavljevi{\'c} 2005; Cooray 2006). Using CLF, one can naturally address the clustering and abundance of galaxies as a function of luminosity which is difficult to study using the basic halo model. Both the basic halo model and the CLF have been successfully applied to local galaxies selected from large spectroscopic surveys such as SDSS and 2dFGRS and distant galaxies selected from DEEP2, COMBO-17 (e.g. Zehavi et al. 2005; Cooray \& Milosavljevi{\'c} 2005; Cooray 2006; Phleps et al. 2006; Coil et al. 2006; Zheng et al. 2007). For example, the halo model can successfully fit the statistically significant deviations from the canonical power-law behaviour in galaxy correlation functions which in turn constrains parameters related to galaxy formation and evolution processes (e.g. the halo mass scale for hosting a central galaxy and satellites). However, so far there are no available halo models of local {\it IRAS} galaxies. Infrared luminosity is a good indicator of star formation activity with is not affected by dust obscuration. Understanding how infrared galaxies populate dark matter haloes could give us important insights on the relation between star formation and environment and the possible mechanisms for triggering / quenching star formation which may preferentially operate on different mass scales (e.g. ram pressure stripping, major merger). 

This paper is organised as follows. In Section 2, we briefly summarise the ingredients in the CLF halo model. Further details can be found in Cooray \& Sheth 2002, Yang et al. 2003, 2005, Cooray 2006. In Section 3, firstly we derive the central galaxy luminosity - halo mass relation and the total galaxy luminosity - halo mass relation using an abundance matching technique. We then constrain the two free parameters left in the CLF model using the 60 $\mu$m LF and the power spectrum of {\it IRAS} galaxies selected from the Point Source Catalogue Redshift (PSCz) catalogue (Saunders et al. 2000). In Section 4, the best-fit CLF model is used to predict the effective galaxy bias as a function of 60 $\mu$m luminosity which agrees well with the clustering measurements of IRAS galaxies in different luminosity bins. In Section 5, we derive the comoving number density of major mergers which is then compared to the measured number density of ultraluminous infrared galaxies (ULIRGs). Discussions and conclusions are presented in Section 6. Throughout the paper, we use a spatially flat $\Lambda$CM cosmology with $\Omega_\Lambda = 0.7$, $\Omega_m=0.3$, the slope of the primordial power spectrum $n_s=1$ and the overall normalisation of the power spectrum $\sigma_8=0.8$. In addition, we build our model at redshift $z=0$ and redshift evolution is not considered.

\section{Conditional luminosity functions (CLF)}

The CLF gives the average number of galaxies that resides in a halo of mass $M$ as a function of luminosity. The CLF is divided into two terms corresponding to central and satellite galaxies
\begin{equation}
\Phi(L|M) = \Phi^{\rm cen}(L|M) + \Phi^{\rm sat}(L|M)
\end{equation}
The first term is the number of central galaxies as a function of luminosity and halo mass which is assumed to obey a log-normal relation,
\begin{equation}
\Phi^{\rm cen}(L|M) = \frac{\Phi(M)}{\sqrt{2\pi}\ln (10) \sigma_{\rm cen} L} {\rm exp} \left[-\frac{\log_{10}[L/L_c(M)]^2}{2\sigma_{\rm cen}} \right].
\end{equation}
The mean luminosity $L_c(M)$ encodes the relation between central galaxy luminosity and halo mass and $\sigma_{\rm cen}$ represents the dispersion in this relation. The second term corresponds to the luminosity distribution (assumed to be a power-law) of satellite galaxies which appear when the total luminosity inside a halo exceeds the central galaxy luminosity,
\begin{equation}
\Phi^{\rm sat}(L|M) = A(M) L^{\gamma}.
\end{equation}
The normalisation $\Phi(M)$ and $A(M)$ are such that $\int \Phi^{\rm cen}(L|M)LdL=L_c(M)$ and $\int \Phi^{\rm sat}(L|M)LdL=L_{\rm tot}(M) - L_c(M)$. The galaxy LF can then be obtained by integrating CLFs over the halo mass function,
\begin{equation}
\Phi(L) = \int dM n(M) \Phi(L|M). 
\end{equation}
Here $n(M)$ represents the mass function of dark matter halos and we use the formalism of Sheth \& Tormen (1999),
\begin{eqnarray}
n(M)dM & = &\frac{\bar{\rho}}{M^2}\nu f(\nu)\left|\frac{d \ln \sigma}{d \ln M}\right| dM\\
\nu f(\nu) &=& 2A \left(1+\frac{1}{\nu'^{2q}}\right)\left(\frac{\nu'^2}{2\pi}\right)^{1/2}\exp \left(-\frac{\nu'^2}{2}\right)
\end{eqnarray}
where $\bar{\rho}$ is the mean matter density of the Universe at $z=0$, $\sigma(M)$ is the linear rms mass fluctuation on mass scale $M$,  $\delta_c$ is the critical overdensity required for collapse at $z=0$,  $\nu=\delta_c/\sigma(M)$ , $\nu' = \sqrt(a)\nu$, $a=0.707$, $q=0.3$ and $A\approx0.322$.

\begin{figure*}\centering
\includegraphics[height=4.0in,width=5.9in]{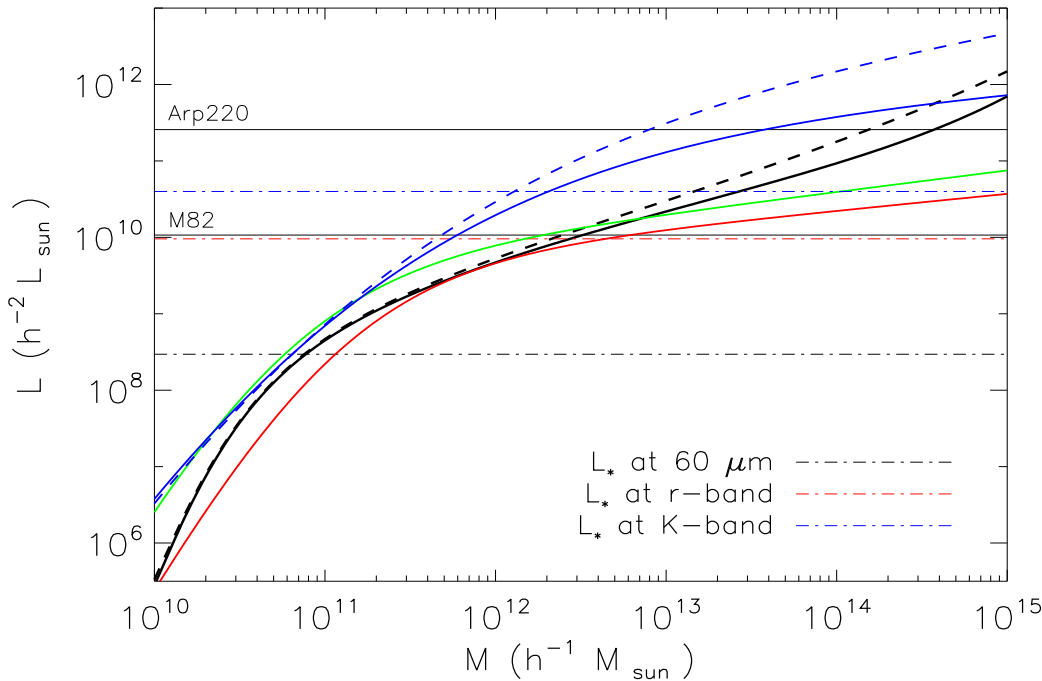}
\caption{The relation between dark matter halo mass and galaxy luminosity. Different solid lines represent the relations between halo mass and central galaxy luminosity at different wavebands (black solid line - {\it IRAS} 60 $\mu$m; red solid lines - SDSS r-band; green solid line - rest B-band; blue solid line - rest K-band). Different dashed lines represent the relation between halo mass and total galaxy luminosity ( black dashed line - {\it IRAS} 60 $\mu$m; blue dashed line - rest K-band. The horizontal solid lines mark the locations above which the 60 $\mu$m LF is dominated M82/Arp220 type starburst. The horizontal dot-dashed lines indicate $L_*$ at different wavebands.}
\label{fig:masslight}
\end{figure*}

Similarly, the galaxy power spectrum can be decomposed into two clustering terms in the halo model
\begin{equation}
P_{\rm gal}(k|L) = P_{\rm 1h}(k|L) + P_{\rm 2h}(k|L).
\end{equation}
The 1-halo contribution comes from galaxies which reside in the same dark matter halo and it can be expressed as
\begin{eqnarray}
P_{\rm 1h}(k|L) & = & \frac{1}{\bar{n}(L)^2} \int n(M) [\Phi^{\rm sat}(L|M)^2 u_{\rm g}(k|M)^2 \nonumber\\
               & + & 2 \Phi^{\rm cen}(L|M) \Phi^{\rm sat}(L|M) u_{\rm g}(k|M)] dM, 
\end{eqnarray} 
where $\bar{n}(L)$ is the mean number density of galaxies as a function of luminosity,
\begin{equation}
\bar{n}(L) = \int dM n(M) \Phi(L|M)
\end{equation}
and $u_{\rm g}(k|M)$ is normalised Fourier transform of the galaxy density distribution within a halo of mass $M$,
\begin{equation}
u_{\rm g}(k|M) = \int_0^{r_{\rm vir}} dr 4\pi r^2 \frac{\sin ~kr}{kr} \frac{\rho_{\rm g}(r|M)}{M}.
\end{equation}
We assume that the galaxy density distribution within the dark matter halos $\rho_{\rm g}(r|M)$ can be described by an NFW profile (Navarro, Frenk \& White 1997) truncated at the virial radius, 
\begin{equation}
M_{\rm vir} = \frac{4}{3} \pi r_{\rm vir}^3 \bar{\rho} \Delta_{\rm vir},
\end{equation}
where the virial overdensity is $\Delta_{\rm vir} \approx 337$ (Bullock et al. 2001). The 2-halo contribution comes from galaxies which reside in separate dark matter halos,
\begin{eqnarray}
P_{\rm 2h}(k|L) &=& P^{\rm lin}(k) \nonumber \\
               &~& \left[\int dM n(M) b(M) \frac{\Phi(L|M)}{\bar{n}(L)} u_{\rm g}(k|M)\right]^2.
\end{eqnarray}
Here $P^{\rm lin}(k)$ is the linear dark matter power spectrum and $b(M)$ is the bias factor for dark matter halos of mass $M$. In this paper, we use the bias factor derived by Sheth, Mo \& Tormen (2001) based on the ellipsoidal collapse model. If one is interested in the galaxy power spectrum above a certain luminosity threshold or over a wide luminosity range, one can simply integrate the CLF $\Phi^{\rm cen}(L|M)$ and $\Phi^{\rm sat}(L|M)$ over luminosity to obtain the halo occupation numbers $N^{\rm cen}(M)$ and $N^{\rm sat}(M)$ and replace the CLF with the halo occupation numbers in Eq. 8, 9 and 12.

\section{Observational Constraints: luminosity function and galaxy power spectrum}

Following Vale \& Ostriker (2004), we derive the relation between the mass of the dark matter halo/subhalo and the luminosity of a galaxy hosted in it using an abundance matching technique. The inputs in this approach are the halo mass function $n(M)$ (Eq. 5 and 6), the mass distribution of subhalos in a halo of a given mass and the galaxy LF. The mass distribution of subhalos in a parent halo of mass $M$ is 
\begin{equation}
N(m|M) dm = A\left(\frac{m}{x\beta M}\right)^{-\alpha} \exp \left ( -\frac{m}{x\beta M}\right) \frac{dm}{x\beta M},
\end{equation}
where $\alpha=1.91$, $\beta=0.39$, $\gamma=0.18$, $x=3$ and $A=\gamma / [\beta\Gamma(2-\alpha)]$ (De Lucia et al. 2004; Weller, Ostriker \& Bode 2005). The total mass in these subhalos is a fraction of the parent halo mass, $x\gamma M$. The global subhalo mass function can thus be obtained by integrating over the halo mass function,.
\begin{equation}
n_{\rm sh}(m) = \int_0^\infty N(m|M) n(M)dM.
\end{equation}
The halo mass - central galaxy luminosity relation can be derived by matching the number density of galaxies above a certain luminosity and the number density of haloes / subhaloes above a certain mass threshold,
\begin{equation}
\int_L^\infty \Phi(L) dL = \int_M^\infty [n(M) + n_{\rm sh}(M)] dM.
\end{equation}
We use the present-epoch 60 $\mu$m LF from Saunders et al. (1990) which is a Gaussian combined with a power law,
\begin{equation}
\varphi(L)=C\left(\frac{L}{L_*}\right)^{1-\alpha} \exp\left[-\frac{1}{2\sigma^2}\log_{10}^2 \left(1+\frac{L}{L_*}\right)\right], 
\label{eqn:Saunders-LF}
\end{equation}
where $C=2.6\times 10^{-2}~h^3~\rm{Mpc}^{-3}$ dex$^{-1}$, $\alpha=1.09$, $\sigma=0.724$, $L_*=10^{8.47}~h^{-2}~L_{\odot}$. Note that $\varphi(L)$ is the LF per decade in luminosity. There are two implicit assumptions in this approach: (1) the luminosity of a galaxy hosted in a halo is an increasing function of the halo mass; (2) Each halo or subhalo can only host up to one galaxy in total. The halo mass - total luminosity relation can be expressed as follows
\begin{equation}
L_{\rm tot}(M) = L(M) + \int_0^\infty L(m) N(m|M) dm,
\end{equation}
which means the total luminosity in a halo of mass $M$ is the sum of its central galaxy luminosity and satellite luminosities in all its subhalos.

Fig.~\ref{fig:masslight} shows the derived mass - luminosity relation for IRAS galaxies selected at 60 $\mu$m. In halos less massive than $10^{12} h^{-1} M_\odot$, the mass - total luminosity relation is essentially the same as the mass - central galaxy luminosity relation. Compared with mass - light relations at other wavebands, e.g. SDSS r-band, rest B-band and rest K-band, the 60 $\mu$m mass - light relation seems to be distinct in the sense that it can not be fitted by the general fitting formula,
\begin{equation}
L_c(M) = L_0 \frac{(M/M_1)^a}{[b+(M/M_1)^{cd}]^{1/d}}.
\end{equation}
This is because the 60 $\mu$m LF does not show a sharp exponential cutoff above $L_*$ and therefore is not well fitted by the Schechter function $\Phi(L) = \Phi_*(L/L_*)^{-\alpha}\exp (-L/L_*)$ (Schechter 1976) which accurately describes the LFs at r-, B- and K-band (e.g. Blanton et al. 2001, 2003; Norberg et al. 2002; Smith et al. 2009). In other words, there are relatively more luminous infrared galaxies than luminous optical galaxies.

The knee of the LF $L_*$ is shown in Fig.~\ref{fig:masslight} for different wavebands. It is immediately obvious that galaxies with luminosities around $L_*$ at r- and K-band reside in halos of similar mass while $L_*$ galaxies selected at 60 $\mu$m are hosted by much less massive halos (roughly by over an order of magnitude). This is consistent with the relative bias between these different tracers of the large-scale structure derived from clustering measurements (e.g. Peacock \& Dodds 1994; Fisher et al. 1994; Mann et al. 1996; Hawkins et al. 2001; Zehavi et al. 2002, 2005; Ma et al. 2009). In Fig.~\ref{fig:masslight}, we also show the luminosity scales $L_{60}=10^{10.0}$ $h^{-2} L_\odot$ and $L_{60}=10^{11.4}$ $h^{-2} L_\odot$ above which the 60 $\mu$m LF is dominated by M82 type (interaction-induced) and Arp220 type (merger-induced) starbursts, respectively (Wang \& Rowan-Robinson 2010). At luminosities below $L_{60}=10^{10.0}$ $h^{-2} L_\odot$, the 60 $\mu$m LF is dominated by cirrus type quiescent star-forming galaxies. In the luminosity range where M82 and Arp220 type starbursts dominate, the rate of change in $L_{60}$ as a function of halo mass $M$ is much larger than the rate of change in $L_{\rm optical}$ or $L_{\rm NIR}$. Since the 60 $\mu$m luminosity can be used as a proxy for star formation rate, it implies that in halos with masses $\gtrsim 10^{12.5}h^{-1}M_\odot$, there are some physical mechanisms which are very efficient in converting cold gas into stars. From numerical simulations, groups have a virial mass of $\sim 10^{13}~h^{-1}~M_\odot$, while clusters have virial masses of $10^{14}$ (poor) - $10^{15}~h^{-1}~M_\odot$ (rich). So, one possibility is that dynamical effects such as gravitational interaction and merging activity in these more massive halos trigger a frantic phase of star formation. Observationally, most local luminous infrared galaxies show signs of interaction or merging activities at various stages (e.g. Sanders \& Mirabel 1996; Clements et al. 1996; Murphy et al. 1996; Farrah et al. 2001; Veilleux, Kim \& Sanders 2002). Numerical simulations have also shown that tidal forces in interacting / merging galaxies are very efficient at creating centrally concentrated gas - fuel for nuclear starburst and active galactic nuclei activity (e.g. Mihos \& Hernquist 1994a,b, 1996; Springel 2000; Cox et al. 2006; Di Matteo et al. 2007). We will return to this point in Section 5.

\begin{figure}\centering
\includegraphics[height=3.0in,width=3.4in]{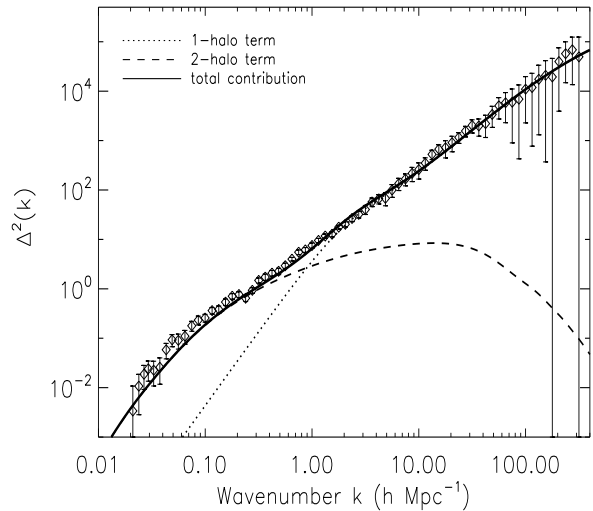}
\caption{The {\it IRAS} PSCz galaxy power spectrum (empty diamonds) measured in Hamilton \& Tegmark (2002) compared with the predicted power spectrum from the best-fit CLF model.}
\label{fig:clustering}
\end{figure}

\begin{figure}\centering
\includegraphics[height=3.0in,width=3.4in]{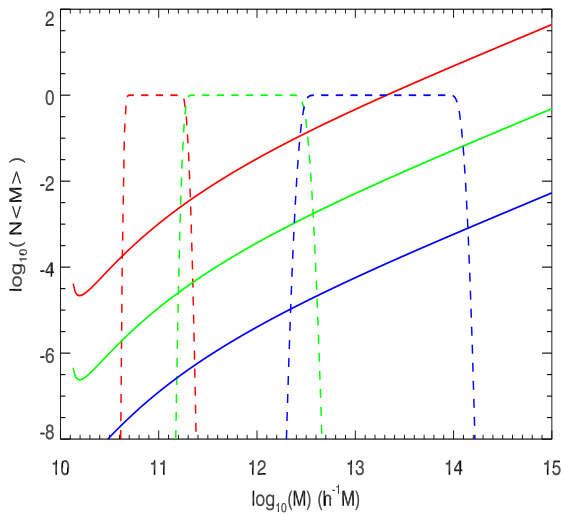}
\caption{The halo occupation numbers as a function of halo mass. The solid lines correspond to satellite galaxies and the dashed lines correspond to central galaxies. The red, green and blue curves represent the halo occupation numbers for galaxies in the luminosity range [10$^8$, 10$^9$], [10$^9$, 10$^{10}$] and [10$^{10}$, 10$^{11}$] $h^{-2}$ $L_\odot$, respectively.}
\label{fig:HOD}
\end{figure}

\begin{figure*}\centering
\includegraphics[height=4.0in,width=5.4in]{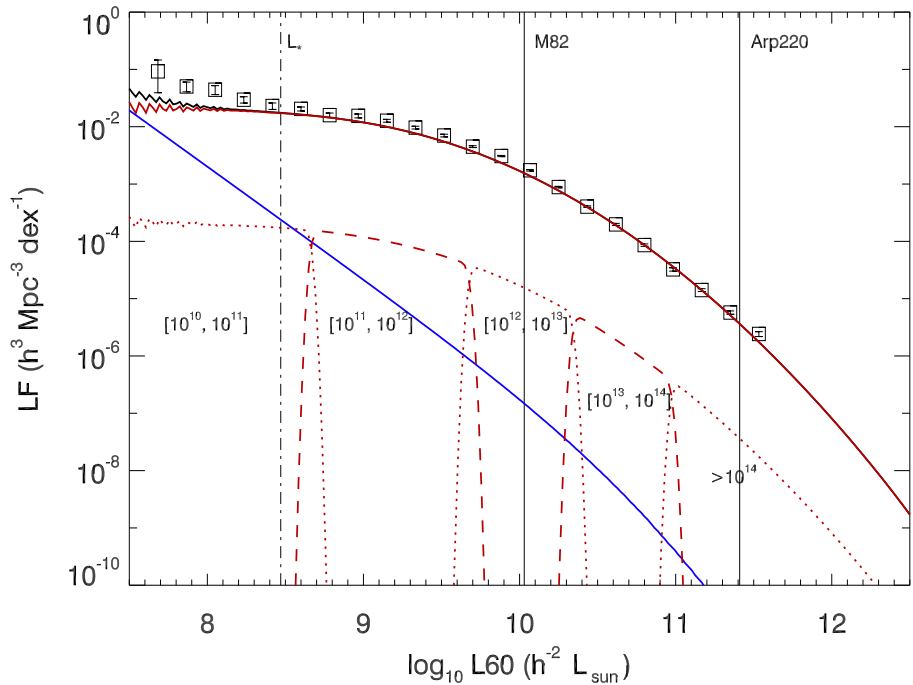}
\caption{The symbols are the 60 $\mu$m luminosity function of {\it IRAS} galaxies derived from a non-parametric maximum likelihood method (Wang \& Rowan-Robinson 2010). The red solid line corresponds to the contribution from central galaxies while the blue solid line corresponds to the contribution from satellites. The black solid lines shows the sum of the two. The red dotted and dashed lines shows the contribution of central galaxies in different halo mass ranges indicated by the numbers shown under each curve. The amplitude of the red dotted and dashed lines have been reduced by a factor of 100 to avoid overcrowding.}
\label{fig:LF60}
\end{figure*}

Now we have derived the relation between central / total galaxy luminosity and halo mass, the only free parameters left in CLF are the power-law slope $\gamma$ in the luminosity distribution of satellite galaxies and the dispersion $\sigma_{\rm cen}$ in the luminosity distribution of central galaxies. We vary both $\sigma_{\rm cen}$ and $\gamma$ over a wide range and search for the best-fit values that minimise $\chi^2=\chi^2(\Phi(L)) + \chi^2(P(k))$. For the LF $\Phi(L)$, we use the non-parametric maximum likelihood estimates from Wang \& Rowan-Robinson (2010). For the galaxy power spectrum $P(k)$, we use the real-space power spectrum of the {\it IRAS} PSCz survey from Hamilton \& Tegmark (2002). We find that the best-fit parameters are $\gamma=-2.96$ and $\sigma_{\rm cen}=0.022$. In Fig.~\ref{fig:clustering}, we compare the {\it IRAS} PSCz power spectrum with the predicted power spectrum from the best-fit model. The 1-halo clustering term is sensitive to $\gamma$ - the power-law slope in the satellite luminosity distribution. The transition from the 2-halo term to the 1-halo term occurs at $k\sim 1~h~Mpc^{-1}$. Fig.~\ref{fig:HOD} shows the halo occupation numbers for central and satellite galaxies as a function of halo mass. At a give halo mass scale, the number of satellites decreases rapidly as $L_{60}$ increases. At a given luminosity scale, the number of satellites increases rapidly as a function of halo mass. It indicates that a significant fraction of low-luminosity IRAS galaxies are satellites hosted in massive dark matter halos. In Fig. ~\ref{fig:LF60}, the predicted LF and the contribution from central and satellite galaxies from the best-fit CLF model are compared to the measured LF. The 60 $\mu$m LF is dominated by central galaxies at all luminosities apart from the faintest luminosity bins ($L_{60}\approx 10^{7.5}~h^{-2}~L_\odot$), similar to the finding at other wavebands (Cooray et al. 2005). The shape of the LF at the bright end is most sensitive to $\sigma_{\rm cen}$. At luminosities below $L_{60} = 10^{10}~h^{-2}~L_\odot$ where the 60 $\mu$m LF is dominated by cirrus type galaxies, the fraction of satellite galaxies increases rapidly as $L_{60}$ decreases. This implies that many satellites in massive halos could have on-going quiescent star formation.

%The bias factor formula
%\begin{eqnarray}
%b(M) &=&1+\frac{1}{\sqrt{a}\delta_c}\left[ \sqrt{a}(a\nu^2) + \sqrt{a}b(a\nu^2)^{1-c} \nonumber\\
%     &-&\frac{(a\nu^2)^2}{ (a\nu^2)^c + b(1-c)(1-c/2)} \right],
%\end{eqnarray}
%where $a=0.707$, $b=0.5$, $c=0.6$.

\section{Testing the best-fit model}

\begin{figure}\centering
\includegraphics[height=3in,width=3.4in]{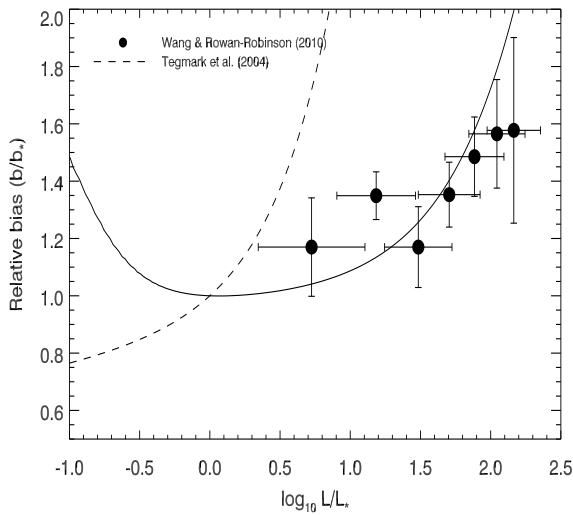}
\caption{The measured effective bias of {\it IRAS} galaxies as a function of 60 $\mu$m luminosity compared with the predicted luminosity dependence from the best-fit CLF halo model. The dashed line represents the relative bias for optical galaxies derived from SDSS power spectrum measurements, $b_{\rm g}(L) / b_{\rm g}(L_*)=0.85 + 0.15L/L_*-0.04(M-M_*)$ (Tegmark et al. 2004). We use $L_*=10^{8.47}~h^{-2}~L_{\odot}$ for IRAS galaxies and $L_*=10^{10}~h^{-2}~L_{\odot}$ for optical galaxies.}
\label{fig:bias}
\end{figure}

Using the real-space galaxy auto-correlation functions, Wang \& Rowan-Robinson (2010) measured the bias factor as a function of 60 $\mu$m luminosity using galaxies from the Imperial IRAS-FSC Redshift Catalogue (IIFSCz; Wang \& Rowan-Robinson 2009). In this section, we test our best-fit CLF model using the luminosity dependence of galaxy clustering. From the best-fit CLF model, we can derive the effective bias factor as a function of luminosity,
\begin{equation}
b_{\rm g}(L) = \int dM n(M) b(M) \frac{\Phi(L|M)}{\bar{n}(L)}.
\end{equation}
In Fig.~\ref{fig:bias}, we plot the predicted bias factor as a function of 60 $\mu$m luminosity. A good agreement ($\chi^2_{\rm red} = 1.1$) is found between the model prediction and the measured bias factors. The dashed line in Fig.~\ref{fig:bias} shows galaxy bias as function of luminosity for optical galaxies derived from measurements of the SDSS power spectrum (Tegmark et al. 2004), $b_{\rm g}(L) / b_{\rm g}(L_*)=0.85+0.15L/L_*-0.04(M-M_*)$. We use $L_*=10^{8.47}~h^{-2}~L_{\odot}$ for IRAS galaxies and $L_*=10^{10}~h^{-2}~L_{\odot}$ for optical galaxies. Compared with the bias derived from optical galaxies, the clustering properties of IRAS galaxies show a rather mild luminosity dependence around $L_*$. This is because IRAS $L_*$ galaxies reside in low mass halos where the linear bias factor does not change rapidly as a function of halo mass (e.g. Sheth \& Tormen 1999; Sheth, Mo \& Tormen 2001; Seljak \& Warren 2004). The rise of galaxy bias towards lower $L_{60}$ is caused by an increasing fraction of low luminosity IRAS galaxies being hosted as satellites in more massive halos. In addition, we do not find any evidence for merger bias which is a tendency of recently merged systems to be more strongly clustered on large scales than typical systems of similar mass (e.g. Furlanetto \& Kamionkowski 2006; Chapman et al. 2009).

\section{ULIRGs and major mergers}

Local ultraluminous infrared galaxies (ULIRGs) with infrared luminosities $L_{\rm IR} \geq 10^{12}~h^{-2}~L_\odot$ are believed to be gas-rich major merger systems which are evolutionally connected with QSOs (e.g. Mihos \& Hernquist 1996; Sanders \& Mirabel 1996; Moorwood 1996; Lonsdale, Farrah \& Smith 2007; Wang et al. 2010). Combining the two Millennium simulation data, Fakhouri O. et al. (2010) derived the following fitting formula for the dimensionless mean merger rate (i.e. the mean number of mergers per halo),
\begin{equation}
\frac{dN_m}{d\xi dz}(M, \xi, z) = A \left( \frac{M}{10^{12}M_\odot}\right)^\alpha \xi^\beta \exp\left[ \left(\frac{\xi}{\tilde{\xi}}\right)^\gamma\right] (1+z)^\eta,
\end{equation}
where $M$ is the descendant halo mass, $\xi$ is the progenitor mass ratio (i.e. $M_1/M_2$ with $M_1 \leq M_2$), $(\alpha, \beta, \gamma, \eta) = (0.133, -1.995, 0.263, 0.0993)$ and $(A, \tilde{\xi})=(0.0104, 9.72\times 10^{-3})$. The total space density of $z\approx0$ major mergers $n_{\rm mm}$ for halos hosting ULIRGs can be calculated as 
\begin{equation}
n_{\rm mm} = \int_{M_{\rm ULIRG}}^\infty \int_{1/3}^1 n(M) \frac{dN_m}{d\xi dz}(M, \xi, z=0) d\xi dM,
\end{equation}
where $M_{\rm ULIRG}$, the average halo mass scale hosting infrared galaxies with $L_{\rm IR}=10^{12}~h^{-2}~L_\odot$, is derived from the mass - light relation in Fig.~\ref{fig:masslight} using an empirical ratio of $L_{\rm IR} / L_{60}=2.2\pm0.2$. Using a volume-limited sample of ULIRGs from the IIFSCz at $z\leq0.1$, we derive the co-moving space density of local ULIRGs to be around $n_{\rm ULIRG} = 2.2\times 10^{-7} ~h^3~{\rm Mpc}^{-3}$. Compared to the theoretical estimate in the same redshift range $n_{\rm mm} = 1.1\times 10^{-6}~h^3~{\rm Mpc}^{-3}$, it implies that the majority of major mergers are not observed as ULIRGs\footnote{There are a few caveats here. Firstly, a halo - halo major merger may not be the same as a galaxy - galaxy major merger which is believed to trigger the ULIRG phase. However, it is not clear which mass ratio definition is the correct one to use, stellar mass, baryonic mass, dynamical mass or halo mass. Secondly, the conversion from the halo - halo merger rate to the galaxy - galaxy merger rate is highly uncertain mainly due to the uncertainties in simulating baryonic physics. For example, by adopting different merger timescales (i.e. the time delay between a halo - halo merger and  the subsequent galaxy - galaxy merger), the predicted galaxy - galaxy merger rate can differ by a factor of $\sim2$. Hopkins et al. (2010) investigated various error sources which in combination can lead to order-of-magnitude variation in the predicted galaxy-galaxy merger rates.}. This could be due to either a large fraction of the major merger events at $z\sim0$ are dry mergers rather than wet mergers or the ultraluminous infrared phase is relatively short-lived so that only a small fraction are observed. There is still a lot of debate on the importance of dry mergers in the formation of massive galaxies. Some numerical studies show that dry mergers dominate the overall merger rate at $z\sim0$ (e.g. Naab et al. 2006; Khochfar \& Burkert 2003, 2005; Khochfar \& Silk 2009). The short duration of the ULIRGs phase is probably more likely to be the cause of the difference between $n_{\rm ULIRG}$ and $n_{\rm mm}$ from a variety of arguments. Stellar population synthesis models show that the UV and optical spectra of ULIRGs are consistent with star formation bursts of $\sim 10^7$ - $10^8$ years old (e.g. Canalizo \& Stockton 2000a,b, 2001; Farrah et al. 2005; Rodr{\'{\i}}guez Zaur{\'{\i}}n et al. 2008, 2009). Similar numbers come from gas depletion time estimates. Assuming the bulk of infrared luminosity arises from star formation, the gas depletion time is $M_{\rm gas}/{\rm SFR} \sim 10^7 - 10^8$ years (e.g. Sanders et al. 1988; Solomon et al. 1997; Downes \& Solomon 1998; Evans et al. 2002; Papadopoulos et al. 2008).

\section{Discussion and conclusion}

We present a halo model of local IRAS galaxies selected at 60 $\mu$m using conditional luminosity functions (CLFs) $\Phi(L|M)$. We match the number of galaxies above a certain luminosity threshold to the number of halos and subhalos above a certain mass threshold to derive the relation between halo mass and the 60 $\mu$m luminosity. The implicit assumptions in this approach are: (1) The mass - light relation is monotonic; (2) every halo / subhalo can only host up to one galaxy. The resultant mass - light relation at 60 $\mu$m is very different from those derived at r, K and B-band at the bright end because the 60 $\mu$m luminosity function (LF) is not well fitted by a Schechter function with a sharp exponential cutoff above $L_*$. The typical halo mass scales hosting $L_*$ galaxies at 60 $\mu$m, r- and K-band are consistent with the measured relative bias between infrared galaxies, optical galaxies and near-infrared selected galaxies. In addition, the rate of change in $L_{60}$ as a function of halo mass is much larger that the rate of change in $L_{\rm optical}$ or $L_{\rm NIR}$, indicating the existence of physical mechanisms in these halos which are efficient in inducing star formation activities. Gravitational interactions and major mergers are likely to be the cause of high star formation rates based on morphological studies of LIRGs and ULIRGs.

Building on the mass -light relation at 60 $\mu$m, we seek the parameters in the CLF model that best fit the 60 $\mu$m LF and the power spectrum of {\it IRAS} PSCz galaxies. In the best-fit model, we find that the majority of IRAS galaxies with $L_{60} \gtrsim 10^{10}~h^{-2}~L_\odot$ are M82 type starbursts which are central galaxies hosted in halos of mass $\gtrsim 10^{12.5}~h^{-1}~M_{\odot}$. The resultant best-fitting model is also tested using the luminosity dependence of galaxy clustering. A good agreement is found between the predicted bias as a function of $L_{60}$ and the clustering measurement of IRAS galaxies in different luminosity bins. Compared to optical galaxies, IRAS galaxies show a mild luminosity-dependent clustering around the characteristic luminosity $L_*$. This is because IRAS $L_*$ galaxies are hosted in low mass halos where the halo bias does not change rapidly with mass. In addition, we do not find evidence for merger bias.

Lastly, we compare the predicted major merger number density with the measured number density of ultraluminous infrared galaxies (ULIRGs). We find that the former is about five times larger than the latter. This could mean that the majority of the major merger events do not lead to ULIRGs. However, this is probably more likely to be explained by the relatively short duration of the ULIRG phase.

In a future paper, we will use CLF halo model to fully model the cross-correlation functions between different types of infrared galaxies and the cross-correlation functions between infrared galaxies from IRAS and optical galaxies from SDSS. This work can also be extended to model the relative distribution between infrared galaxies detected by Herschel and other classes of galaxies in the distance Universe (Amblard \& Cooray 2007). It would also be interesting to compare the CLF model constrained by observational data with predictions from semi-analytic modelling. 

\section*{ACKNOWLEDGEMENTS}
L. Wang and S. Oliver are supported by UK's Science and Technology Facilities Council grant ST/F002858/1. A. Cooray acknowledges support from NSF CAREER AST-0645427. L. Wang thanks David Seery, Antony Lewis, Duncan Farrah, Peter Thomas, Anthony Smith and Christopher Short for useful discussions. This work has made use of the public conditional luminosity function code by A. Cooray and M. Milosavljevi{\'c}.

\end{document}